\documentclass[runningheads]{llncs}
\usepackage[T1]{fontenc}
\usepackage{graphicx}

\newcommand{\anonymize}[1]{#1} 
\newcommand{\osf}[1]{\footnote{See Files section: \url{https://osf.io/b2y9n/?view\_only=bf2d045f35ae453faa225d92017d48cb}, last accessed May 2025}} 

\begin{document}
%
\title{When motivation can be more than a message: designing agents to boost physical activity}
\titlerunning{Designing agents to boost physical activity}
%

\author{Alessandro Silacci\inst{1,2}\orcidID{0000-0001-8121-3013} \and
Maurizio Caon\inst{2}\orcidID{0000-0003-4050-4214} \and
Mauro Cherubini\inst{1}\orcidID{0000-0002-1860-6110}}
\authorrunning{\anonymize{Silacci et al.}}
%

\institute{Persuasive Technology Lab, University of Lausanne, Switzerland
\email{\{alessandro.silacci, mauro.cherubini\}@unil.ch} \and
School of Management of Fribourg, HES-SO University of Applied Sciences and Arts of Western Switzerland, Switzerland
\email{\{alessandro.silacci, maurizio.caon\}@hes-so.ch}}
\maketitle              
\begin{abstract}
Virtual agents are commonly used in physical activity interventions to support behavior change, often taking the role of coaches that deliver encouragement and feedback. While effective for compliance, this role typically lacks relational depth. This pilot study explores how such agents might be perceived not just as instructors, but as co-participants: entities that appear to exert effort alongside users. Drawing on thematic analysis of semi-structured interviews with 12 participants from a prior physical activity intervention, we examine how users interpret and evaluate agent effort in social comparison contexts. Our findings reveal a recurring tension between perceived performance and authenticity. Participants valued social features when they believed others were genuinely trying. In contrast, ambiguous or implausible activity levels undermined trust and motivation. Many participants expressed skepticism toward virtual agents unless their actions reflected visible effort or were grounded in relatable human benchmarks. Based on these insights, we propose early design directions for fostering co-experienced exertion in agents, including behavioral cues, narrative grounding, and personalized performance. These insights contribute to the design of more engaging, socially resonant agents capable of supporting co-experienced physical activity.
\end{abstract}
\section{Introduction}
\label{sec:introduction}
    
In Human-Computer Interaction (HCI), intelligent agents have long been used to support behavior change, especially in physical activity contexts where they often take the form of virtual coaches, delivering encouragement, tracking performance, and offering feedback~\cite{bennettRolePatientAttachment2011,bickmoreEstablishingComputerPatient2005,fadhilAssistiveConversationalAgent2019,kocielnikReflectionCompanionConversational2018,maherPhysicalActivityDiet2020}. While this role is effective for promoting compliance~\cite{fadhilAssistiveConversationalAgent2019,maherPhysicalActivityDiet2020}, it remains limited in depth, focusing on instruction rather than fostering richer, more relational forms of interaction~\cite{bennettRolePatientAttachment2011,bickmoreEstablishingComputerPatient2005,kocielnikReflectionCompanionConversational2018}.

\par Agents are uniquely positioned to influence motivation through not just informational support but also social presence, a factor known to sustain long-term engagement~\cite{renExploringCooperativeFitness2018,stragierComputerMediatedSocialSupport2018}. However, current designs rarely leverage this potential to create a sense of shared experience, particularly in the expression of effort.

\par Recent advancements in Large Language Models (LLMs) have expanded the capabilities of virtual agents, enabling more fluid, context-sensitive, and empathetic interactions~\cite{jorkeGPTCoachLLMBasedPhysical2025}. However, in the realm of physical activity interventions, these agents currently remain confined to traditional coaching models providing textual support, missing opportunities to foster deeper engagement through shared experiences and co-participation.

\par In contrast, agents in video games are often designed as opponents, companions, or mirrors of the player -- roles that foster emotional connection, narrative engagement, and meaningful social dynamics. Concepts like Simulated Exercising Peers (SEPs) have begun to explore similar dynamics in physical activity contexts~\cite{silacciNavigatingDesignSimulated2025}, but it remains unclear how users perceive an agent’s effort, or what makes that effort feel believable and motivating.

\par A key challenge is how agents can meaningfully convey exertion. Unlike humans, they do not visibly struggle or tire, and users may not naturally attribute effort to them. Identifying the cues, behaviors, and design elements that make an agent’s effort feel believable, motivating, and shared with the user is a key challenge for future agent-based interventions.

\par The aim of this pilot study is to explore how virtual agents might be perceived as co-participants in physical activity interventions -- engaging not only as coaches, but as entities that appear to share the user’s effort. We are particularly interested in how users interpret agent behavior as indicative of exertion or collaboration. To guide this investigation, we ask: \textit{How do users perceive and evaluate virtual agents’ effort in the context of social physical activity interventions?} Through a series of qualitative interviews, we examine the cues, expectations, and conditions that shape these perceptions, and consider how such insights might inform the design of more socially resonant and relational agent interactions.

\section{Background}
\label{sec:background}
Sustaining physical activity remains a persistent challenge, despite interventions grounded in motivational theories such as SDT and SCT, which emphasize relatedness, autonomy, and social modeling~\cite{banduraHealthPromotionPerspective1998,deciWhatWhyGoal2000}. One of the strategy has been to embed social mechanisms such as support, feedback, and peer comparison, into interventions aiming to enhance motivation through interpersonal dynamics~\cite{stragierComputerMediatedSocialSupport2018}.

\par However, coordinating group-based activity poses practical challenges, often sustained engagement and synchronized schedules. To address these limitations, researchers have turned to virtual agents, including conversational and embodied forms, as scalable alternatives that simulate social presence. These agents can personalize feedback and interaction timing while maintaining a sense of continuity and responsiveness. 

\par Yet, systems that incorporate social comparison present a double-edged sword: while motivating for some, they can discourage others when competition feels unfair or unbalanced~\cite{estabrooksGroupDynamicsPhysical2012,molleeEffectivenessUpwardDownward2016}. This tension is particularly relevant for Simulated Exercising Peers (SEPs), agents designed to support human teammates through social support and physical effort simulation~\cite{silacciNavigatingDesignSimulated2025}. In this work, we explore how agents might go beyond comparison and coaching by fostering a perception of effort, enabling more engaging and cooperative interactions.

\subsection{Effort, embodiment, and perception}
\label{sub:background_perception_hci}
While social comparison can motivate physical activity, it also risks unfair competition and reduced self-confidence, particularly when performance disparities are pronounced~\cite{estabrooksGroupDynamicsPhysical2012,molleeEffectivenessUpwardDownward2016}. These risks extend to agent-based systems~\cite{zhouExploringUserExperience2023}, including SEPs~\cite{silacciNavigatingDesignSimulated2025}, which must balance challenge and fairness. To avoid replicating the pitfalls of human competition, agent design should shift from emphasizing performance outcomes to supporting perceived effort -- both in the user and in the agent.

\par Cognitive science and psychology emphasize that effort is not simply a measure of physical strain, but a subjective experience -- a felt cost associated with continuing an action~\cite{bermudezWhatFeelingEffort2025}. This internal sense of effort helps guide decisions about persistence, disengagement, and intensity. Crucially, motivation depends not only on how difficult a task is in objective terms, but on how demanding it feels. This logic extends to social contexts, where perceiving effort in others can influence commitment and engagement in cooperative activities.

\par Empirical studies support this view. Users invest more effort when they believe a partner is exerting themselves~\cite{chennellsEffortPerformanceCooperative2018}, and subtle cues of agent commitment -- even without performance data -- enhance perseverance~\cite{szekelyPerceptionRobotPartners2019}. Responsiveness and socially supportive behaviors further increase users’ feelings of closeness and engagement~\cite{zhouExploringUserExperience2023}. Yet current physical activity agents rarely communicate effort in socially meaningful ways. This presents a key opportunity: to design agents not just to act, but to be seen as trying -- fostering a sense of co-experienced exertion between human and machine.

\subsection{Roles of agents in physical activity}
\label{sub:background_agents_pa}
In physical activity interventions, virtual agents are most commonly cast in the role of coaches -- providing encouragement, guiding goal setting, and offering educational support to help users build healthier habits~\cite{fadhilAssistiveConversationalAgent2019,jorkeGPTCoachLLMBasedPhysical2025,kocielnikReflectionCompanionConversational2018,maherPhysicalActivityDiet2020}. These agents do more than dispense advice; by engaging users in sustained dialogue, they can foster relational bonds that support behavior change over time~\cite{bickmoreEstablishingComputerPatient2005}. These agents can increase user adherence by facilitating conversation and building strong relational ties~\cite{bickmoreEstablishingMaintainingLongterm2005}. As these relationships develop, users often come to see the agent as a collaborative partner -- a dynamic described in therapeutic contexts as a working alliance, built on shared goals, trust, and mutual engagement~\cite{bennettRolePatientAttachment2011}. 

\par Although coaching agents are often preferred for their structure and expertise~\cite{griffithsExerciseSocialRobots2018}, companions provide alternative forms of support. These agents, often imagined as animals, cyborgs, or abstract creatures, use informal communication and can appear more emotionally responsive~\cite{silacciNavigatingDesignSimulated2025}. Their design enables users to influence the agent’s state, such as its health~\cite{kniestedtLittleFitnessDragon2016} or mood~\cite{linFishnStepsEncouragingPhysical2006}, promoting a more reciprocal dynamic.

\par The role an agent plays has significant implications for interaction design. Coaches tend to function as authoritative figures, directing tasks and goals~\cite{mohanDesigningAIHealth2020}, which can streamline structure but may limit relational flexibility~\cite{salmanIdentifyingWhichRelational2023}. Companion agents, in contrast, support more emotionally resonant and socially dynamic relationships~\cite{kniestedtLittleFitnessDragon2016,linFishnStepsEncouragingPhysical2006}. Their informality allows for increased social presence and vulnerability, making them especially promising for scenarios involving shared struggle or exertion. Critically, while coaches often monitor user effort, companion agents may better support the perception of agent effort -- enabling richer, more immersive experiences of co-experienced physical activity~\cite{silacciNavigatingDesignSimulated2025}. 

\par While companion agents may help convey a sense of shared effort, it remains unclear what cues lead users to perceive such exertion -- a gap this pilot study explores.

\section{Methods}
\label{sec:methods}
To investigate how users perceive virtual agents’ physical effort, we conducted semi-structured interviews with 12 participants from an inter-group competition experiment promoting physical activity. The six-month study involved a prototype app where participants viewed their own and teammates’ daily step counts on a leaderboard. Each team competed weekly against another, with access to both intra-team and opponent leaderboards.

\par We collected qualitative data through post-experiment interviews, focusing on how participants interpreted teammate behavior and how they envisioned virtual agents fitting into such settings. Our goal was not to assess performance outcomes, but to identify early themes in how effort is perceived -- and which design elements might support or hinder that perception.
\subsection{Recruitment}
\label{sub:interviews_methods_recruitment}
Participants for the interviews were selected from a larger ongoing experiment at the \anonymize{University of Lausanne}, specifically those exposed to features designed to provide social connectedness (e.g., a team leaderboard and user profiles). As the intervention used cumulative weekly step counts as a proxy for physical performance, we aimed to recruit participants representing a range of activity levels: below, at, and above the median step count. Recruitment was conducted via email from the [redacted] participant pool, which includes over $8,000$ volunteers, primarily university students. A total of $96$ participants were eligible for inclusion.

\subsection{Participants}
\label{sub:interviews_methods_participants}
We selected $37$ eligible participants based on their willingness to participate, availability and their fluency in French, of which $13$ agreed to participate. One was excluded due to consent form issues, resulting in a final sample of $N = 12$ (8 female, $M_{age} = 23\ years$). All participants had previously engaged in an intervention that tracked weekly step counts as a performance measure, and put them in an inter-group competition setup. The median weekly count was $31259.5$; among interviewees, $N_{higher}$ = 3 had values above the median and $N_{lower} = 9$ below. A detailed selection notebook is available in our Open Science Framework (OSF) repository\osf.

\subsection{Procedure}
\label{sub:interviews_methods_procedure}

We conducted one-hour semi-structured interviews with participants previously involved in a mobile intervention at \anonymize{University of Lausanne}, each of whom received a compensation of USD 27 \anonymize{(CHF 25)}. The first part of the interview focused on two app features: a team leaderboard that displayed weekly step counts to elicit social comparison~\cite{estabrooksGroupDynamicsPhysical2012}, and user profiles with avatars, pseudonyms, hobbies, and preferred sports. We asked how these features shaped engagement, teammate comparisons, and users’ sense of relatedness within the app’s coopetitive structure (intra-team collaboration, inter-group competition).

\par In the second part, we explored perceptions of virtual agents. Participants were asked how such agents might complement existing features, what traits would make them credible or relatable, and how they would feel about collaborating or competing with AI teammates. Special attention was given to how agent effort might be perceived and expressed, and whether such agents could meaningfully participate in social dynamics around physical activity.

\subsection{Analysis}
\label{sub:interviews_methods_analysis}
Content generated through the interviews was analyzed using thematic analysis~\cite{braunUsingThematicAnalysis2006} to examine how participants interpreted social dynamics, perceived effort, and the potential role of agents in physical activity contexts. Our analysis was guided by the research question: \textit{How do users perceive and evaluate virtual agents’ effort in the context of social physical activity interventions?}

\par We used a primarily inductive, semantic coding approach. After jointly reviewing a subset of transcripts, two researchers collaboratively developed an initial set of $22$ open codes and  $8$ axial codes grounded in the data. These codes were iteratively refined through regular meetings as new patterns and nuances emerged during full dataset analysis. The complete codebook and thematic structure are available in our OSF repository\osf.

\subsection{Ethical Consideration}
\label{sub:interviews_methods_ethical}
The Institutional Review Board (IRB) of the \anonymize{University of Lausanne} approved our study. Interviews were conducted online and participants were provided with a link to the online meeting and could join as anonymous guests. We ensured anonymity of the data by replacing the used names of the interviewees with anonymous identifiers (e.g., P1, P2, etc.). 

\section{Results}
\label{sec:results}
We identified three main themes in the qualitative data, reflecting how participants perceived social dynamics, effort, and credibility within the intervention.

\subsection{Competition as a potential barrier}
\label{sub:results_competition}

Participants frequently engaged in social comparison, often aspiring to match high-performing teammates. However, this same dynamic could become demotivating when perceived as unfair or unrepresentative. Technical issues (e.g., lost data) and contextual disruptions (e.g., exams) made some feel their effort was invisible. One participant, after losing progress due to app failure, remarked, \textit{``It’s just that when I saw that I was losing everything [because of the app tracking issues] I was there: Well I, mean I walk quite a bit, so…''} -- reflecting both frustration and resignation.

\par Perceptions of inactive or implausibly high-performing teammates further eroded trust. Some questioned whether certain users were even real, speculating that a teammate \textit{``may have been a computer.''} This ambiguity led to what we interpret as a breakdown in social grounding -- participants could not reliably assess others’ effort or contextualize their own.

\par To cope with underperformance, some participants withdrew from the app, hid it, or avoided checking the leaderboard. These avoidance behaviors suggest a self-protective response to upward comparison, consistent with prior work showing that such comparisons can backfire when users lack control or credibility cues~\cite{estabrooksGroupDynamicsPhysical2012,molleeEffectivenessUpwardDownward2016}.

\par Overall, while competition initially drove engagement, it also risked triggering guilt and disengagement. These findings underscore the need for systems to represent effort fairly and transparently to sustain motivation.

\subsection{Credibility and the search for explanations}
\label{sub:results_credibility}
Participants often assessed teammates’ profiles not for social connection, but to determine whether their activity levels were believable. They scanned details like age, hobbies, and listed sports to evaluate whether performance on the leaderboard made sense. As one participant put it, seeing a teammate with \textit{``doing 20,000 steps per day''} led them to assume that person \textit{``must have been a runner.''} Such inferences illustrate how users constructed plausible narratives to make comparative data feel credible.

\par This interpretive work reflects a deeper need to validate the fairness of social comparisons. When profile cues matched expectations, users were more likely to trust others’ effort. When data seemed implausible or lacked context, participants grew skeptical -- in some cases questioning whether certain users were even real. This credibility gap weakened trust in the system and diminished the motivational value of the leaderboard.

\par In short, participants used personal plausibility as a stand-in for perceived effort, anchoring trust in social data through coherence between identity and performance.

\subsection{Skepticism toward agent effort and authenticity}
\label{sub:results_effort}
A recurring source of frustration among participants was the lack of concrete information about their teammates, which created uncertainty about whether those users were real. Some described the experience as \textit{``dehumanized''} or \textit{``very anonymous,''} noting that pseudonyms and avatars alone were not enough to establish genuine social presence. This was particularly problematic when performances were extreme -- either very high or unusually low -- which participants found difficult to contextualize without additional background.

This skepticism extended to the idea of integrating virtual agents. Many participants expressed discomfort with the notion that teammates might be computer-controlled, stating that it would undermine the sense of authentic competition. As one participant put it, \textit{``the competition is between the people,''} suggesting that human agency was essential to the leaderboard’s motivational effect. Others noted that knowing an agent was \textit{``just a program''} would render its performance meaningless, as it wouldn’t reflect real physical effort -- just \textit{``a number that is useless.''}

Despite this skepticism, several participants proposed ways that agents could still contribute meaningfully. One common idea was to use the agent as a daily step target or motivational benchmark. As one described, it could serve as \textit{``a motor for motivation,''} offering inspiration rather than competition. Others suggested that grounding the agent’s behavior in the identity or performance of a real person -- such as an athlete or public figure -- could make its actions feel more credible and aspirational.

\section{Discussion and conclusion}
\label{sec:discussion_and_conclusion}
Our findings point to a recurring tension between performance and perceived effort. Participants valued social features when they felt that their teammates’ activity reflected real exertion -- a dynamic that virtual agents currently struggle to convey. If an agent’s behavior is interpreted as effortless or artificial, its motivational credibility diminishes. This highlights a key challenge for the design of agents in physical activity contexts: not just how they act, but how they appear to try.

\par Insights from game design may help illuminate this challenge. In video games, agents -- whether companions or opponents -- are often designed to exhibit adaptive difficulty, emotional expression, or visible struggle~\cite{chowandaPlayingSocialEmotional2016,emmerichImGladYou2018}. These features are not only aesthetic but functional: they signal effort, fallibility, and investment, which foster engagement and emotional connection. For example, when game agents exhibit emotional expressiveness and vulnerability -- such as fatigue or personality-driven responses -- players report greater immersion and social connection~\cite{chowandaPlayingSocialEmotional2016}.

\par Conversely, when agents are perceived as infallible or overpowered, the relationship can break. The case of AlphaGo is illustrative: despite its technical mastery, human players found its style demotivating, in part because its moves appeared opaque and devoid of struggle~\cite{egri-nagyGameNotGo2020}. In our pilot study, participants expressed a similar unease when they imagined competing against virtual teammates whose step counts were \textit{``just a number''} -- lacking context, backstory, or any sign of exertion.

\par These parallels suggest that designers of physical activity agents might draw on game-based conventions to reinforce perceptions of effort and co-experience. Agents could, for instance, visibly slow down when the user is inactive, display fatigue when achieving high step counts, or adapt their pacing to mirror the user’s progress. By signaling effort through expressive behavior, narrative grounding, or performance variability, agents might move beyond role-based interactions (e.g., coach) toward more emotionally resonant partnerships.  Based on our findings, we suggest the following preliminary design directions to inform future agent development:

\begin{enumerate}
    \item \textbf{Make agent effort visible} Participants responded strongly to perceived authenticity. Agents should display signs of exertion -- such as slowed movement, visual fatigue, or verbal cues -- to signal they are ``working'' too. This can strengthen believability and mutual challenge.
	\item \textbf{Use human-like performance benchmarks} To avoid seeming unfair or artificially perfect, agent behavior should reflect realistic human standards (e.g., average step counts or variability), helping users interpret the agent’s effort as credible.
    \item \textbf{Personalize for co-engagement} Agents can foster connection by adjusting to the user’s pace -- slowing down when they do, or progressing in sync. Such alignment supports the illusion of shared effort and builds relational depth.
\end{enumerate}

\par \textbf{Limitations.} While these insights offer promising directions for design, several limitations should be noted in interpreting the findings. As an exploratory pilot, our findings are based on a small and relatively homogeneous sample of university students. Data was drawn from semi-structured interviews, relying on participant imagination rather than live interaction with functioning agents. We also did not triangulate interview responses with behavioral or usage data, which could have added further depth and validation. Future research should involve interactive prototypes and broader participant samples to better understand how perceptions of agent effort play out in real-world settings.

\par Moving forward, the challenge is to bridge imagined potential with embodied design -- creating agents that can visibly participate in effort, adapt to users, and foster sustained motivation through shared experience.

\begin{credits}
\subsubsection{\ackname}
The authors acknowledge financial support from the \anonymize{HEC} Faculty at the \anonymize{University of Lausanne}, which funded participant compensation.

\subsubsection{\discintname}
The authors have no competing interests to declare that are relevant to the content of this article.
\end{credits}

\newpage
\bibliographystyle{splncs04}
\bibliography{phd_alessandro}
\end{document}